\documentclass[12pt]{article}
\usepackage{amsmath}

\newcommand{\ie}{{\em i.e.\/\ }}

\newtheorem{definition}{Definition}

\begin{document}
\title{Optimal Alphabetic Ternary Trees}
\author{ J. David Morgenthaler\thanks{Google, Mountain View, CA 94303. 
	 Email:~{\tt jdm@google.com}. } 
\and 	 
	 T. C. Hu\thanks{Department of Computer Science and Engineering,
	 University of California, San Diego, CA 92093. 
	 Email:~{\tt hu@cs.ucsd.edu}. }
}
%\date{}   % TODO: uncomment to remove date for final submission
\maketitle

\begin{abstract}
We give a new algorithm to construct optimal alphabetic ternary trees, where
every internal node has at most three children. This algorithm generalizes the
classic Hu-Tucker algorithm, though the overall computational complexity has yet
to be determined.

\end{abstract}

\section{Introduction}

The optimal ternary alphabetic tree problem generalizes the well-studied binary
case to consider trees with internal nodes having at most three children. This
problem was mentioned in algorithm textbooks~\cite{hu02, knuth73}, but has
remained unsolved for 40 years..

First, let us briefly review the optimal binary alphabetic tree problem.
Given a sequence of square nodes with weights $\{w_1, w_2,\ldots, w_n\}$.
Find the binary alphabetic tree with the least total path length.
The problem was solved by the Hu-Tucker algorithm (see~\cite{hu02, hu71}).
There are two kinds of nodes, where every circular node has two children,
and every square node has no children.  A pair of nodes is called compatible
if there is no square node between the pair. The weight of the parent node
is the sum of weights of its children.

The Hu-Tucker algorithm has three phases:
 
\subsection*{Phase I: Combination}

Start with a sequence of $n$ square nodes. Keep combining the compatible pair
of nodes with the least weight, creating a new circular node.  In case of a tie,
select the pair with the leftmost position. Keep combining until only one
circular node (the root) remains.

\subsection*{Phase II: Assignment}

Mark the level of each square node with its distance from the root in Phase I.
Thus we obtain a sequence of level numbers $\{l_1, l_2,\ldots, l_n\}$ corresponding
to the $n$ square nodes in Phase I.

\subsection*{Phase III: Reconstruction}

Start with the original sequence of $n$ square nodes, and the corresponding level
sequence.  Find the adjacent pair of square nodes whose level numbers, say $q$,
are maximum among all adjacent pairs. In the case of a tie, select the leftmost
pair. Replace the adjacent pair in the sequence with a new parent node with level
$q-1$. Continue until the sequence consists of a single node with level $0$. The
resulting tree is the optimal binary alphabetic tree.
\vspace{1em}
\par

If we apply the Hu-Tucker Algorithm to the sequence of four square nodes with
weights $\{ 4, 2, 3, 4\}$, then we have the following results.

\subsubsection*{Phase I: Combination}
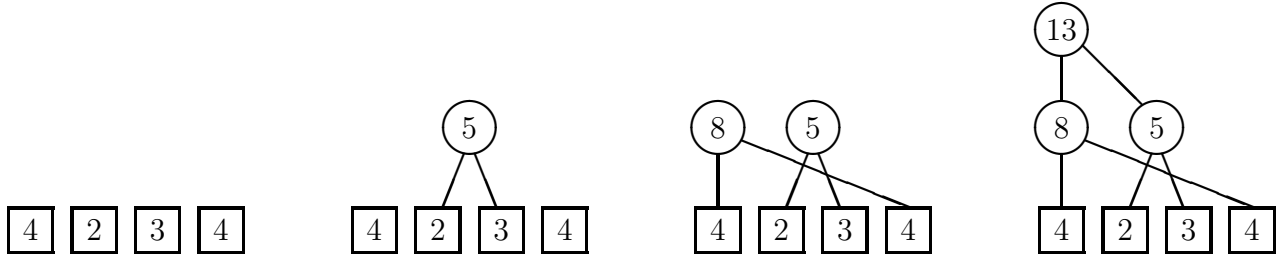
\newsavebox{\Seleven}
\newsavebox{\Sten}
\newsavebox{\Smten}
\newsavebox{\Ssix}
\newsavebox{\Sfive}
\newsavebox{\Sfour}
\newsavebox{\Sthree}
\newsavebox{\Stwo}
\newsavebox{\Sone}
\savebox{\Smten}(20,16){-10}
\savebox{\Seleven}(16,16){11}
\savebox{\Sten}(16,16){10}
\savebox{\Ssix}(16,16){6}
\savebox{\Sfive}(16,16){5}
\savebox{\Sfour}(16,16){4}
\savebox{\Sthree}(16,16){3}
\savebox{\Stwo}(16,16){2}
\savebox{\Sone}(16,16){1}
\begin{center}
\begin{picture}(480, 100)
\thicklines
\put(0,8){\frame{\usebox{\Sfour}}}
\put(24,8){\frame{\usebox{\Stwo}}}
\put(48,8){\frame{\usebox{\Sthree}}}
\put(72,8){\frame{\usebox{\Sfour}}}
\put(130,8){\frame{\usebox{\Sfour}}}
\put(154,8){\frame{\usebox{\Stwo}}}
\put(174,55){\circle{18}}
\put(171,51){5}
\put(176,45){\line(2,-5){8.2}}
\put(172,45){\line(-2,-5){8.2}}
\put(178,8){\frame{\usebox{\Sthree}}}
\put(202,8){\frame{\usebox{\Sfour}}}
\put(260,8){\frame{\usebox{\Sfour}}}
\put(268,55){\circle{18}}
\put(265,51){8}
\put(268,45.1){\line(0,-1){20.3}}
\put(277,50){\line(5,-2){63.5}}
\put(284,8){\frame{\usebox{\Stwo}}}
\put(304,55){\circle{18}}
\put(301,51){5}
\put(306,45){\line(2,-5){8.2}}
\put(302,45){\line(-2,-5){8.2}}
\put(308,8){\frame{\usebox{\Sthree}}}
\put(332,8){\frame{\usebox{\Sfour}}}
\put(390,8){\frame{\usebox{\Sfour}}}
\put(398,55){\circle{18}}
\put(395,51){8}
\put(398,45){\line(0,-1){20.3}}
\put(407,50){\line(5,-2){63.5}}
\put(398,92){\circle{18}}
\put(392,88){13}
\put(398,82){\line(0,-1){17}}
\put(406,86){\line(1,-1){22.3}}
\put(414,8){\frame{\usebox{\Stwo}}}
\put(434,55){\circle{18}}
\put(431,51){5}
\put(436,45){\line(2,-5){8.2}}
\put(432,45){\line(-2,-5){8.2}}
\put(438,8){\frame{\usebox{\Sthree}}}
\put(462,8){\frame{\usebox{\Sfour}}}
\end{picture}
\end{center}

\subsubsection*{Phase II: Assignment}
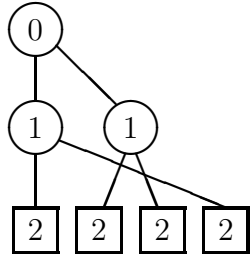
\begin{center}
\begin{picture}(95,100)
\thicklines
\put(0,8){\frame{\usebox{\Stwo}}}
\put(8,55){\circle{18}}
\put(5,51){1}
\put(8,45){\line(0,-1){20.3}}
\put(17,50){\line(5,-2){63.5}}
\put(8,92){\circle{18}}
\put(5,88){0}
\put(8,82){\line(0,-1){17}}
\put(16,86){\line(1,-1){22.3}}
\put(24,8){\frame{\usebox{\Stwo}}}
\put(44,55){\circle{18}}
\put(41,51){1}
\put(46,45){\line(2,-5){8.2}}
\put(42,45){\line(-2,-5){8.2}}
\put(48,8){\frame{\usebox{\Stwo}}}
\put(72,8){\frame{\usebox{\Stwo}}}
\end{picture}
\end{center}

\subsubsection*{Phase III: Reconstruction}
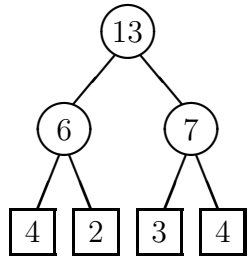
\begin{center}
\begin{picture}(95,100)
\thicklines
\put(0,8){\frame{\usebox{\Sfour}}}
\put(20,55){\circle{18}}
\put(17,51){6}
\put(22,45){\line(2,-5){8.2}}
\put(18,45){\line(-2,-5){8.2}}
\put(24,8){\frame{\usebox{\Stwo}}}
\put(44,92){\circle{18}}
\put(38,88){13}
\put(49,83){\line(5,-6){15.5}}
\put(39,83){\line(-5,-6){15.5}}
\put(48,8){\frame{\usebox{\Sthree}}}
\put(68,55){\circle{18}}
\put(65,51){7}
\put(70,45){\line(2,-5){8.2}}
\put(66,45){\line(-2,-5){8.2}}
\put(72,8){\frame{\usebox{\Sfour}}}
\end{picture}
\end{center}

\begin{definition}
The combination of two compatible nodes with one or more circular nodes between
them is called a cross-over combination.  For example, the creation of a
circular node with weight $8$ by the combination of two square nodes each with 
weight $4$ is called a cross-over.
\end{definition}

\begin{definition}
A forest without any cross-over combinations is called a $k$-sum forest if there
are $k$ combinations. For example, in the Phase III reconstruction above, the 
resulting tree is a $3$-sum forest.
\end{definition}

\noindent To prove correctness, we need two things:
\begin{enumerate}\vspace{-1 mm}
	\item We can always convert any tree with cross-overs to a forest with
             the same number of sums without and cross-over.
        \item The final binary alphabetic tree has the minimum cost.
\end{enumerate}

{\bf Proof} of (1).  Let us assume that our first cross-over has three circular nodes
between the pair $X \; Y$

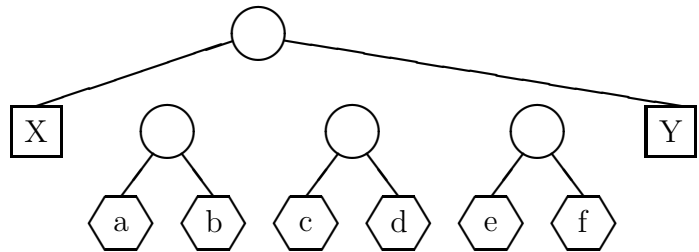
\newsavebox{\hex}
\newsavebox{\hexabcdef}
\newsavebox{\binarycircle}

\begin{center}
\begin{picture}(200, 95)
\thicklines
\savebox{\hex}(16,16)[bl] {
\put(4,10){\line(3,5){6}}
\put(22,0){\line(3,5){6}}
\put(4,10){\line(3,-5){6}}
\put(22,20){\line(3,-5){6}}
\put(10,20){\line(1,0){12}}
\put(10,0){\line(1,0){12}}}
\savebox{\hexabcdef}(220,20)[bl] {
\put(0,0){\usebox{\hex}}
\put(13,7.3){a}
\put(35,0){\usebox{\hex}}
\put(48,7){b}
\put(70,0){\usebox{\hex}}
\put(83.2,7.3){c}
\put(105,0){\usebox{\hex}}
\put(118,7){d}
\put(140,0){\usebox{\hex}}
\put(153.2,7.3){e}
\put(175,0){\usebox{\hex}}
\put(188.8,7){f}}
\put(0,0){\usebox{\hexabcdef}}
\savebox{\binarycircle}(40,40)[bl] {
\put(33.5,45){\circle{18}}
\put(16.2,20.5){\line(3,4){12}}
\put(51,20.5){\line(-3,4){12}}}
\put(68,82){\circle{18}}
\put(-16,54.55){\line(3,1){74.2}}
\put(228,54.55){\line(-6,1){150}}
\put(-25,36){\framebox(18,18){X}}
\put(215,36){\framebox(18,18){Y}}
\put(0,0){\usebox{\binarycircle}}
\put(70,0){\usebox{\binarycircle}}
\put(140,0){\usebox{\binarycircle}}
\end{picture}
\end{center}

\noindent where hexagons denote either a circular or a square node.  Then we can always
do the four combinations as shown below.

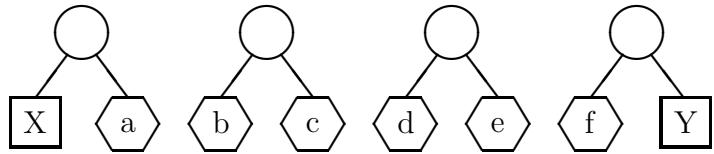
\begin{center}
\begin{picture}(200, 65)
\thicklines
\savebox{\hex}(16,16)[bl] {
\put(4,10){\line(3,5){6}}
\put(22,0){\line(3,5){6}}
\put(4,10){\line(3,-5){6}}
\put(22,20){\line(3,-5){6}}
\put(10,20){\line(1,0){12}}
\put(10,0){\line(1,0){12}}}
\savebox{\hexabcdef}(220,20)[bl] {
\put(0,0){\usebox{\hex}}
\put(13,7.3){a}
\put(35,0){\usebox{\hex}}
\put(48,7){b}
\put(70,0){\usebox{\hex}}
\put(83.2,7.3){c}
\put(105,0){\usebox{\hex}}
\put(118,7){d}
\put(140,0){\usebox{\hex}}
\put(153.2,7.3){e}
\put(175,0){\usebox{\hex}}
\put(188.8,7){f}}
\put(0,0){\usebox{\hexabcdef}}
\savebox{\binarycircle}(40,40)[bl] {
\put(33.5,45){\circle{18}}
\put(16.2,20.5){\line(3,4){12}}
\put(51,20.5){\line(-3,4){12}}}
\put(-28,2){\framebox(18,18){X}}
\put(218,2){\framebox(18,18){Y}}
\put(-35,0){\usebox{\binarycircle}}
\put(35,0){\usebox{\binarycircle}}
\put(105,0){\usebox{\binarycircle}}
\put(175,0){\usebox{\binarycircle}}
\end{picture}
\end{center}

Where in the last two figures, the total path lengths are the same, and both
are created with $4$ sums, and with the same alphabetic order.  This proves
that the final tree is alphabetic.

To prove that the final binary tree is optimal, \ie of minimal cost, we can use
induction on the sums during the algorithm.  After the first $k$ combinations,
we can apply phases II and III to obtain an optimal $k$-sum forest.  This leads
to an interpretation of the weight of each circular node as the incremental cost 
of moving from an optimal $(k-1)$-sum forest to an optimal $k$-sum forest. 

\section{Optimal Ternary Alphabetic Trees}

Given a sequence of square nodes with weights $\{w_1, w_2,\ldots, w_n\}$ where
every circular node has $t$ children $(t \le 3)$, and every square node has no
children.  The optimal alphabetic ternary tree has the least total path length
for the given sequence of weights. 

This problem was mentioned by Knuth~\cite{knuth73}, p.\ 451, problem 40.  
He presented a simple five node sequence $\{1, 1, 100, 1, 1\}$ demonstrating
the difficulty in applying a straight-forward generalization of the Hu-Tucker
algorithm to triples. The sequence shows that optimal trees may include interior
nodes with two as well as three children, as seen in its optimal ternary tree below.

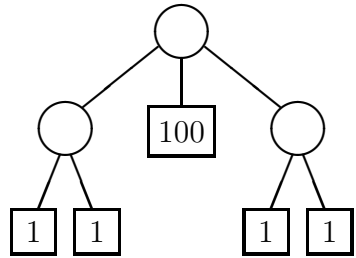
\begin{center}
\begin{picture}(120,110)
\thicklines
\put(0,8){\frame{\usebox{\Sone}}}
\put(20,55){\circle{18}}
\put(22,45){\line(2,-5){8.2}}
\put(18,45){\line(-2,-5){8.2}}
\put(24,8){\frame{\usebox{\Sone}}}
\put(64,92){\circle{18}}
\put(52,45){\framebox(24,18){100}}
\put(55.1,86){\line(-5,-4){28.4}}
\put(64,82){\line(0,-1){19}}
\put(72.9,86){\line(5,-4){28.4}}
\put(88,8){\frame{\usebox{\Sone}}}
\put(108,55){\circle{18}}
\put(110,45){\line(2,-5){8.2}}
\put(106,45){\line(-2,-5){8.2}}
\put(112,8){\frame{\usebox{\Sone}}}
\end{picture}
\end{center}
\vspace{-1 mm}

Knuth's example requires us to reconsider the idea of a {\em Permanent Circular
Node}~\cite{hu96} from the binary case and additionally explore the impact of 
odd or even cardinality of a sub-sequence on the combination phase. To begin
with, clearly a two square node initial sequence requires a binary circular node.
 
\subsection{Binary Circular Nodes}

Let our initial sequence be a set of square nodes with monotonically increasing
weights $\{a, b, \ldots\}$. If we have an odd number of nodes, our optimal
ternary tree can consist of internal nodes with exactly three children each. It is
easy to see that a tree that instead contains any binary nodes must cost more. 

On the other hand, if the number of nodes in the initial sequence is even, we
must have at least one binary combination. The following example shows that we 
should do that binary combination in the beginning.  For four square nodes of 
weights $a, b, c, d$, where  $a < b < c < d$
 
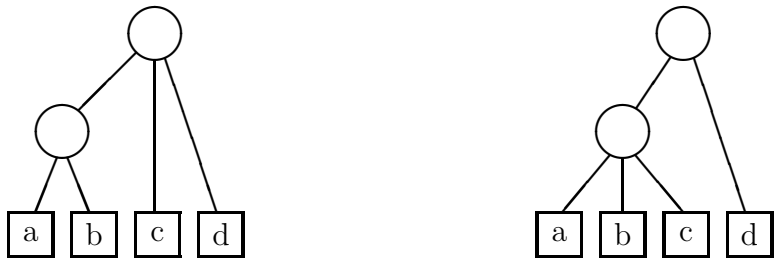
\begin{center}
\begin{picture}(300,100)
\thicklines
\put(0,8){\framebox(16,16){a}}
\put(20,55){\circle{18}}
\put(22,45){\line(2,-5){8.2}}
\put(18,45){\line(-2,-5){8.2}}
\put(24,8){\framebox(16,16){b}}
\put(48,8){\framebox(16,16){c}}
\put(55,92){\circle{18}}
\put(48,85){\line(-1,-1){21.7}}
\put(55,82){\line(0,-1){57.8}}
\put(59,82.7){\line(1,-3){19.4}}
\put(72,8){\framebox(16,16){d}}
\put(200,8){\framebox(16,16){a}}
\put(232,55){\circle{18}}
\put(237,46){\line(5,-6){17.8}}
\put(232,45){\line(0,-1){20.3}}
\put(227,46){\line(-5,-6){17.8}}
\put(224,8){\framebox(16,16){b}}
\put(255,92){\circle{18}}
\put(250,83){\line(-2,-3){12.8}}
\put(259,82.7){\line(1,-3){19.4}}
\put(248,8){\framebox(16,16){c}}
\put(272,8){\framebox(16,16){d}}
\end{picture}
\end{center}

On the left, the total cost $= 2a + 2b + c + d$, while on the right, the total
cost $= 2a + 2b + 2c + d$.

\vspace{4 mm}

If the optimal tree contains binary nodes, they are at the bottom combining two
squares. Specifically, they will be minimum adjacent pair(s). Knuth's sequence
has two such pairs, each meeting the {\em Permanent Circular
Node} criteria we described for optimal binary alphabetic trees~\cite{hu96}. We
generalize this idea to the ternary case. 

\subsection{Generalized Permanent Circular Nodes}

Consider a subsequence of adjacent square nodes with weights $w_i, \ldots, w_j$
where their combined weight is smaller than either neighbor $w_{i-1}$ or $w_{j+1}$
$$w_{i-1} > w_i + \ldots + w_j < w_{j+1}$$

For convenince, we can add nodes $w_0$ and $w_{n+1}$ of infinite weight
at either end of the intial sequence to handle edge cases. These added nodes
never combine, but are simply an algorithmic convenience.

Generalizing Gilbert and Moore's Theorem 5~\cite{gilbert59} to the ternary
case is straightforward, but results in two possibilities:

\begin{enumerate}\vspace{-1 mm}
	\item All nodes in the subsequence combine to form a single circle before
	 combination with either larger adjacent node.
	\item Two nodes, rather than a single one, are available to combine with one neighbor.
\end{enumerate}

The decision as to which of the two possibilites to use will be determined at the next
level, by the evenness or oddness of the cardinality of that larger subproblem. 
Taken to the top level, the complete sequence can also be viewed as a `permanent' 
circular node between the two added infinite-weight nodes. For this final case, 
of course, we ignore the two node possibility.

Applying this idea to Knuth's example, the five node sequence $\{1, 1, 100, 1, 1\}$
contains only three nodes considering the generalized permanent nodes as units:
$\{(1, 1), 100, (1, 1)\}$.  Three is odd, and so forms a ternary circle, implying that the
two permanent nodes each combine to form a single circle. We can increase the
number of squares in each permanent node to see the idea in operation.

Adding a third square to the initial permanent node, $\{1, 1, 1, 100, 1, 1\}$ 
results in a similar situation as before, just turning the first pair into a triple:
$\{(1, 1, 1), 100, (1, 1)\}$. If we add a forth square to that group, only those
four nodes are affected, and the four node sub-problem will provide a single
permanent cirular to the next level as before, consisting of one pair as we saw
in the previous section.

Now consider instead adding another heavy node to Knuth's example sequence:
$$\{1, 1, 100, 100, 1, 1\}$$. The cardinality of the larger problem is even, and
one of the permanent circular nodes must provide two nodes, rather than just a
single one:  $\{(1), (1), 100, 100, (1, 1)\}$.  The resulting five node sequence
has optimal ternary tree $\{((1, 1, 100), 100, (1, 1))\}$, where each internal node
with children $a, b$ appears between parentheses $(a, b)$.

\subsection{Pure Ternary Case}

As a simplifying constraint, we now consider the case where $t=3$.  That is,
every circle must have exactly three children. For this case, the initial sequence
has an odd number of nodes, and we avoid any hierarchal sub-problems by
disallowing permanent circular nodes. That is, for every adjacent pair
$w_i, w_{i+1}$, either $$w_{i-1} \le w_i + w_{i+1}$$ or $$w_i + w_{i+1} \ge w_{i+2}$$
For this purpose only, we consider added negative infinite weight nodes at the
ends of the sequence.

This problem was explored by by Hu and Shing~\cite{hu02},  pp.\ 196-7. Their
counter-example to a naive application of the Hu-Tucker algorithm
uses the initial weight sequence  $\{6, 6, 1, 10, 1, 6, 6\}$. Approaching this problem 
from the viewpoint of consecutive optimal $k$-sum forests, for $k \in \{1,2,3\}$:

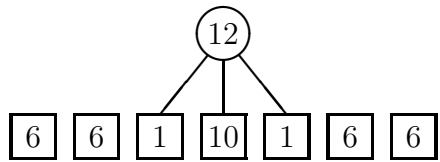
\begin{center}
\begin{picture}(168, 70)
\thicklines
\put(0,8){\frame{\usebox{\Ssix}}}
\put(24,8){\frame{\usebox{\Ssix}}}
\put(48,8){\frame{\usebox{\Sone}}}
\put(72,8){\frame{\usebox{\Sten}}}
\put(96,8){\frame{\usebox{\Sone}}}
\put(120,8){\frame{\usebox{\Ssix}}}
\put(144,8){\frame{\usebox{\Ssix}}}
\put(80,55){\circle{18}}
\put(74,51){12}
\put(74,46.8){\line(-4,-5){17.8}}
\put(80,45){\line(0,-1){20.3}}
\put(86,46.8){\line(4,-5){17.8}}
\end{picture}
\end{center}

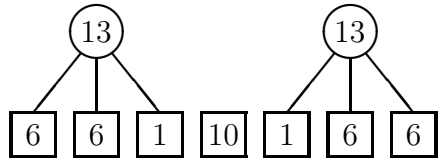
\begin{center}
\begin{picture}(168, 80)
\thicklines
\put(0,8){\frame{\usebox{\Ssix}}}
\put(24,8){\frame{\usebox{\Ssix}}}
\put(48,8){\frame{\usebox{\Sone}}}
\put(72,8){\frame{\usebox{\Sten}}}
\put(96,8){\frame{\usebox{\Sone}}}
\put(120,8){\frame{\usebox{\Ssix}}}
\put(144,8){\frame{\usebox{\Ssix}}}
\put(32,55){\circle{18}}
\put(26,51){13}
\put(26,46.8){\line(-4,-5){17.8}}
\put(32,45){\line(0,-1){20.3}}
\put(38,46.8){\line(4,-5){17.8}}
\put(128,55){\circle{18}}
\put(122,51){13}
\put(122,46.8){\line(-4,-5){17.8}}
\put(128,45){\line(0,-1){20.3}}
\put(134,46.8){\line(4,-5){17.8}}
\end{picture}
\end{center}

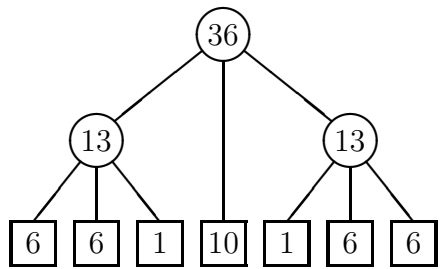
\begin{center}
\begin{picture}(168, 120)
\thicklines
\put(0,8){\frame{\usebox{\Ssix}}}
\put(24,8){\frame{\usebox{\Ssix}}}
\put(48,8){\frame{\usebox{\Sone}}}
\put(72,8){\frame{\usebox{\Sten}}}
\put(96,8){\frame{\usebox{\Sone}}}
\put(120,8){\frame{\usebox{\Ssix}}}
\put(144,8){\frame{\usebox{\Ssix}}}
\put(32,55){\circle{18}}
\put(26,51){13}
\put(26,46.8){\line(-4,-5){17.8}}
\put(32,45){\line(0,-1){20.3}}
\put(38,46.8){\line(4,-5){17.8}}
\put(128,55){\circle{18}}
\put(122,51){13}
\put(122,46.8){\line(-4,-5){17.8}}
\put(128,45){\line(0,-1){20.3}}
\put(134,46.8){\line(4,-5){17.8}}
\put(80,95){\circle{18}}
\put(74,91){36}
\put(72,88.8){\line(-5,-4){33}}
\put(80,85){\line(0,-1){60.3}}
\put(88,88.8){\line(5,-4){33}}
\end{picture}
\end{center}

Note that the associated level numbers of the seven square nodes, at each step,
are

$$0,\;0,\;1,\;1,\;1,\;0,\;0$$
$$1,\;1,\;1,\;0,\;1,\;1,\;1$$
$$2,\;2,\;2,\;1,\;2,\;2,\;2$$

\noindent where the central square node with weight 10 first has level 1, which then
decreases to zero, and finally increases back to level 1 again. This differs from the
binary case where levels increase monotonically as more combinations are made.
Any Hu-Tucker type ternary algorithm must allow for such level reduction during the
combination phase, since the combination phase can always be stopped after $k$
combinations, yielding the level forest for the optimal $k$-sum alphabetic forest.

\section{Ternary Algorithm}

Our new algorithm uses the same three phases as the Hu-Tucker binary algorithm.
Only the combination phase operates differently, as we must generalize the idea
of compatible pair to compatible triple and define a new concept, the accordion. 
A set of three nodes is called a compatible triple if there is no square nodes
between the three nodes. 

\subsection{Accordions}

The main new concept in our algorithm is the accordion.  An accordion forms the
central node of a combining triple, and consists of alternating positive and
negative weight nodes. The outer nodes of an accordion always have positive 
weight, with the smallest accordion consisting of a single node. In the example
of sequence $\{6, 6, 1, 10, 1, 6, 6\}$, the first accordion is the central square
node with weight 10. 

The next combination in the algorithm involves a negative weight, the previously
combined central node of weight 10, this time with weight $-10$.  The accordion
consists of the three central nodes, with weight $(+6, -10, +6) = 2$, giving the
new circle the weight $6 + 2 + 6 = 14$.

Negative weight nodes are previously combined square nodes at level 1.  After
this combination, any negative weight nodes reappear in the sequence as square
nodes, in their original position. Continuing the example, the sequence after 
each phase I combination appears below.

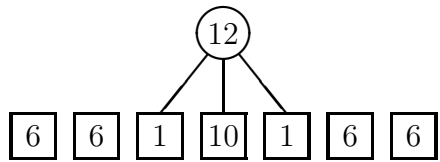
\begin{center}
\begin{picture}(168, 80)
\thicklines
\put(0,8){\frame{\usebox{\Ssix}}}
\put(24,8){\frame{\usebox{\Ssix}}}
\put(48,8){\frame{\usebox{\Sone}}}
\put(72,8){\frame{\usebox{\Sten}}}
\put(96,8){\frame{\usebox{\Sone}}}
\put(120,8){\frame{\usebox{\Ssix}}}
\put(144,8){\frame{\usebox{\Ssix}}}
\put(80,55){\circle{18}}
\put(74,51){12}
\put(74,46.8){\line(-4,-5){17.8}}
\put(80,45){\line(0,-1){20.3}}
\put(86,46.8){\line(4,-5){17.8}}
\end{picture}
\end{center}

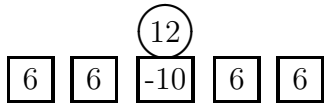
\begin{center}
\begin{picture}(168, 80)
\thicklines
\put(21,8){\frame{\usebox{\Ssix}}}
\put(45,8){\frame{\usebox{\Ssix}}}
\put(70,8){\frame{\usebox{\Smten}}}
\put(99,8){\frame{\usebox{\Ssix}}}
\put(123,8){\frame{\usebox{\Ssix}}}
\put(80,34){\circle{18}}
\put(74,30){12}
\end{picture}
\end{center}

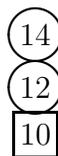
\begin{center}
\begin{picture}(168, 80)
\thicklines
\put(72,8){\frame{\usebox{\Sten}}}
\put(80,34){\circle{18}}
\put(74,30){12}
\put(80,54){\circle{18}}
\put(74,50){14}
\end{picture}
\end{center}

The final combination creates a circle of weight 36 with the three remaining
nodes as children. Note that the square with weight 10 has participated in each
combination, twice with positive weight, and once with negative weight.

A square child of a circle in the current sequence is available with a negative 
weight if that node is the central child of a top-level internal node in the 
unique optimal $k$-sum forest obtained from the current sequence. Reappearing 
squares are positioned below the circle with which they were
associated. Once a level 1 node associated with a particular circle has appeared
as negative weight node in a combination, it cannot be used again.  However, 
the reappearing square node may later combine into a different circle, and again
be used as a negative component of a later accordion.  Once its
circular parent combines, a square loses its availability. Thus, only squares 
at level 2 need be considered.

During phase II, each occurrence of a negative weight in the phase I tree
contributes a negative level number, which when combined with the positive
occurrences of the same square yields the overall level for that square in the
phase III reconstruction.

The proof of correctness and optimality of this algorithm is similar to the
binary case, by induction on the $k$-sum forest created by ending phase I after
the first $k$ combinations and proceeding with phases II and III.

\section{General Case}

Putting the two ideas together yields the general optimal ternary tree algorithm.
First, identify permanent circular nodes with a linear scan of the input. Each 
such node is a subproblem with two solutions - an optimal ternary tree with a
single root node, or an optimal forest with two roots. Permanent circular nodes
may be hierarchical, and the procedure is repeated up the hierarchy until all
remaining squares are almost uniform, that is no remaining pairs have smaller
combined weight than both their adjacent neighbors.

\section{Example}

Consider initial sequence $\{5, 5, 6, 6, 1, 10, 1, 11, 1, 10, 1, 6, 6, 5, 5\}$. 
After the first two combinations, we have a straight forward $2$-sum forest
with total weight 24:

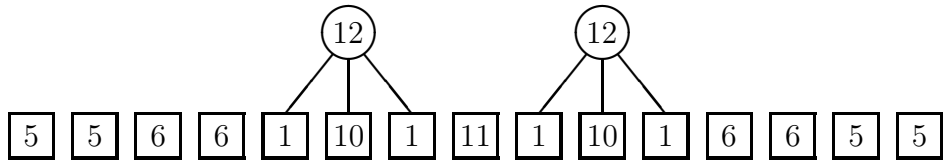
\begin{center}
\begin{picture}(360, 80)
\thicklines
\put(0,8){\frame{\usebox{\Sfive}}}
\put(24,8){\frame{\usebox{\Sfive}}}
\put(48,8){\frame{\usebox{\Ssix}}}
\put(72,8){\frame{\usebox{\Ssix}}}
\put(96,8){\frame{\usebox{\Sone}}}
\put(120,8){\frame{\usebox{\Sten}}}
\put(144,8){\frame{\usebox{\Sone}}}
\put(168,8){\frame{\usebox{\Seleven}}}
\put(192,8){\frame{\usebox{\Sone}}}
\put(216,8){\frame{\usebox{\Sten}}}
\put(240,8){\frame{\usebox{\Sone}}}
\put(264,8){\frame{\usebox{\Ssix}}}
\put(288,8){\frame{\usebox{\Ssix}}}
\put(312,8){\frame{\usebox{\Sfive}}}
\put(336,8){\frame{\usebox{\Sfive}}}
\put(128,55){\circle{18}}
\put(122,51){12}
\put(122,46.8){\line(-4,-5){17.8}}
\put(128,45){\line(0,-1){20.3}}
\put(134,46.8){\line(4,-5){17.8}}
\put(224,55){\circle{18}}
\put(218,51){12}
\put(218,46.8){\line(-4,-5){17.8}}
\put(224,45){\line(0,-1){20.3}}
\put(230,46.8){\line(4,-5){17.8}}
\end{picture}
\end{center}

\noindent For the next Phase I combination, we find a large accordion 
$(+6, -10, +11, -10, +6) = 3$, which participates in minimum triple 
$(6, (+6, -10, +11, -10, +6), 6)$ of weight 15, yielding optimal $3$-sum forest
of total weight $24 + 15 = 39$:

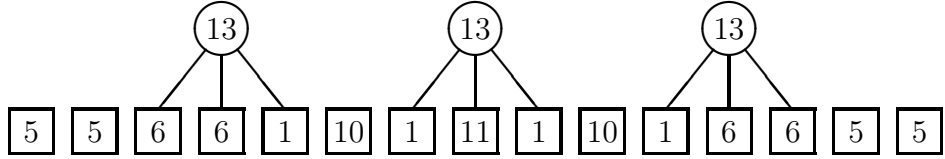
\begin{center}
\begin{picture}(360, 80)
\thicklines
\put(0,8){\frame{\usebox{\Sfive}}}
\put(24,8){\frame{\usebox{\Sfive}}}
\put(48,8){\frame{\usebox{\Ssix}}}
\put(72,8){\frame{\usebox{\Ssix}}}
\put(96,8){\frame{\usebox{\Sone}}}
\put(120,8){\frame{\usebox{\Sten}}}
\put(144,8){\frame{\usebox{\Sone}}}
\put(168,8){\frame{\usebox{\Seleven}}}
\put(192,8){\frame{\usebox{\Sone}}}
\put(216,8){\frame{\usebox{\Sten}}}
\put(240,8){\frame{\usebox{\Sone}}}
\put(264,8){\frame{\usebox{\Ssix}}}
\put(288,8){\frame{\usebox{\Ssix}}}
\put(312,8){\frame{\usebox{\Sfive}}}
\put(336,8){\frame{\usebox{\Sfive}}}
\put(80,55){\circle{18}}
\put(74,51){13}
\put(74,46.8){\line(-4,-5){17.8}}
\put(80,45){\line(0,-1){20.3}}
\put(86,46.8){\line(4,-5){17.8}}
\put(176,55){\circle{18}}
\put(170,51){13}
\put(170,46.8){\line(-4,-5){17.8}}
\put(176,45){\line(0,-1){20.3}}
\put(182,46.8){\line(4,-5){17.8}}
\put(272,55){\circle{18}}
\put(266,51){13}
\put(266,46.8){\line(-4,-5){17.8}}
\put(272,45){\line(0,-1){20.3}}
\put(278,46.8){\line(4,-5){17.8}}
\end{picture}
\end{center}

\noindent An even larger accordion forms the central node of the forth triple combination, 
$$(5, (+5, -6, +10, -11, +10, -6, +5), 5)$$
of weight 17, yielding an optimal $4$-sum forest with total weight $39 + 17 = 56$:

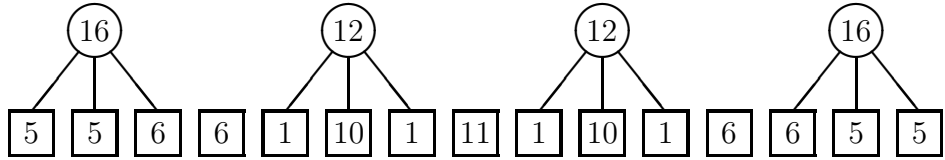
\begin{center}
\begin{picture}(360, 80)
\thicklines
\put(0,8){\frame{\usebox{\Sfive}}}
\put(24,8){\frame{\usebox{\Sfive}}}
\put(48,8){\frame{\usebox{\Ssix}}}
\put(72,8){\frame{\usebox{\Ssix}}}
\put(96,8){\frame{\usebox{\Sone}}}
\put(120,8){\frame{\usebox{\Sten}}}
\put(144,8){\frame{\usebox{\Sone}}}
\put(168,8){\frame{\usebox{\Seleven}}}
\put(192,8){\frame{\usebox{\Sone}}}
\put(216,8){\frame{\usebox{\Sten}}}
\put(240,8){\frame{\usebox{\Sone}}}
\put(264,8){\frame{\usebox{\Ssix}}}
\put(288,8){\frame{\usebox{\Ssix}}}
\put(312,8){\frame{\usebox{\Sfive}}}
\put(336,8){\frame{\usebox{\Sfive}}}
\put(32,55){\circle{18}}
\put(26,51){16}
\put(26,46.8){\line(-4,-5){17.8}}
\put(32,45){\line(0,-1){20.3}}
\put(38,46.8){\line(4,-5){17.8}}
\put(128,55){\circle{18}}
\put(122,51){12}
\put(122,46.8){\line(-4,-5){17.8}}
\put(128,45){\line(0,-1){20.3}}
\put(134,46.8){\line(4,-5){17.8}}
\put(224,55){\circle{18}}
\put(218,51){12}
\put(218,46.8){\line(-4,-5){17.8}}
\put(224,45){\line(0,-1){20.3}}
\put(230,46.8){\line(4,-5){17.8}}
\put(320,55){\circle{18}}
\put(314,51){16}
\put(314,46.8){\line(-4,-5){17.8}}
\put(320,45){\line(0,-1){20.3}}
\put(326,46.8){\line(4,-5){17.8}}
\end{picture}
\end{center}

\noindent At this point, the fifth combination is the three square triple $(6, 11, 6)$, 
for an optimal $5$-sum forest with all nodes at level 1, with weight 
$56 + 23 = 79$. Now our sequence consists of only circular nodes, which 
we place in a queue: $(12, 12, 15, 17, 23)$. Two more combinations - 
$12 + 12 + 15 = 39$ and $17 + 23 + 39 = 79$ completes Phase I giving a total cost
of $79 + 39 + 79 = 197$. 

For each combination, we can also show the current level of each square using 
underscores and overscores. We denote a combination of positive 
weight nodes with an underscore, while reappearing squares of negative weight 
receive an overscore. Each underscore represents an increase in level, while each 
overscore is a decrease. Using this scheme, the Phase I combinations above look
like this:

$$ 5 \;\;\;\; 5 \;\;\;\; 6 \;\;\;\; 6 \;\;\;\; \underline{1\;\;10\;\;1} \;\;\;\; 11 \;\;\;\; 
\underline{1\;\;10\;\;1} \;\;\;\; 6 \;\;\;\; 6  \;\;\;\; 5 \;\;\;\; 5$$

$$ 5 \;\;\;\; 5 \;\;\;\; \underline{6 \;\;\;\; 6 \;\;\;\; 1\;\;\overline{10}\;\;1
\;\;\;\; 11 \;\;\;\; 1\;\;\overline{10}\;\;1 \;\;\;\; 6 \;\;\;\; 6}  \;\;\;\; 5 \;\;\;\; 5$$

$$ \underline{5 \;\;\;\; 5 \;\;\;\; 6 \;\;\;\; \overline{6} \;\;\;\; 1\;\;
\underline{\overline{10}}\;\; 1 \;\;\;\; \overline{11} \;\;\;\; 1\;\;
\underline{\overline{10}}\;\;1 \;\;\;\; \overline{6} \;\;\;\; 6  \;\;\;\; 5 \;\;\;\; 5}$$

$$ \underline{5 \;\;\;\; 5 \;\;\;\; 6 \;\;\;\; \underline{\overline{6}} \;\;\;\; 1\;\;
\underline{\overline{10}}\;\; 1 \;\;\;\; \underline{\overline{11}} \;\;\;\; 1\;\;
\underline{\overline{10}}\;\;1 \;\;\;\; \underline{\overline{6}} \;\;\;\; 6  \;\;\;\; 5 \;\;\;\; 5}$$

$$ \underline{5 \;\;\;\; 5 \;\;\;\; \underline{6 \;\;\;\; \underline{\overline{6}} \;\;\;\; 1\;\;
\underline{\overline{\overline{10}}}\;\; 1 \;\;\;\; \underline{\overline{11}} \;\;\;\; 1\;\;
\underline{\overline{\overline{10}}}\;\;1 \;\;\;\; \underline{\overline{6}} \;\;\;\; 6}  \;\;\;\; 5 \;\;\;\; 5}$$

$$  \underline{\underline{5 \;\;\;\; 5 \;\;\;\; \underline{6 \;\;\;\; \underline{\underline{\overline{\overline{6}}}} \;\;\;\; 1\;\;
\underline{\underline{\overline{\overline{\overline{10}}}}}\;\; 1 \;\;\;\; \underline{\underline{\overline{\overline{11}}}} \;\;\;\; 1\;\;
\underline{\underline{\overline{\overline{\overline{10}}}}}\;\;1 \;\;\;\; \underline{\underline{\overline{\overline{6}}}} \;\;\;\; 6} \;\;\;\; 5 \;\;\;\; 5}}$$

\vspace{.15in}
\noindent Phase II thus yields level numbers:
$$2,\;2,\;3,\;3,\;3,\;2,\;3,\;3,\;3,\;2,\;3,\;3,\;3,\;2,\;2$$

\section{Conclusion and Future Work}

We have extented the Hu-Tucker algorithm to the ternary case, where each
internal node has at most three children.  The ternary case is more complex than
the binary alphabetic tree problem, and future work incudes determination of the 
overall complexity of the solution presented here. There is additional work of 
tracking two possibilities for each permanent circular node and computing the 
oddness or evenness of each subproblem. Accordians must also be tracked, 
and their existance adds work to determining the next phase I combination.

The question of whether a Hu-Tucker type level tree algorithm can be further 
extended to handle $t > 3$ children remains open.

\bibliographystyle{abbrv}
\bibliography{alpha}

\begin{thebibliography}{1}

\bibitem{gilbert59}
E.~N. Gilbert and E.~F. Moore.
\newblock Variable length binary encodings.
\newblock {\em Bell System Technical Journal}, 38:933--968, 1959.

\bibitem{hu96}
T.~C. Hu and J.~D. Morgenthaler.
\newblock Optimum alphabetic binary trees.
\newblock In {\em Combinatorics and Computer Science: 8th Franco-Japanese and
  4th Franco-Chinese Conference}, Lecture Notes in Computer Science, volume
  1120, pages 234--243. Springer-Verlag, 1996.

\bibitem{hu02}
T.~C. Hu and M.~T. Shing.
\newblock {\em Combinatorial Algorithms, Second Edition}.
\newblock Dover, 2002.

\bibitem{hu71}
T.~C. Hu and A.~C. Tucker.
\newblock Optimal computer search trees and variable-length alphabetic codes.
\newblock {\em SIAM Journal on Applied Mathematics}, 21(4):514--532, 1971.

\bibitem{knuth73}
D.~E. Knuth.
\newblock {\em The Art of Computer Programming, Volume III: Sorting and
  Searching}.
\newblock Addison-Wesley, 1973.

\end{thebibliography}

\end{document}